\documentclass[final,5p,times,twocolumn]{elsarticle}

\usepackage{amssymb}
\usepackage{amsmath}

\biboptions{sort&compress}
\biboptions{}

\usepackage{float}
\usepackage{graphicx}
\usepackage{dcolumn}
\usepackage{bm}
\usepackage{xcolor}
\usepackage[breaklinks,colorlinks,citecolor=blue,urlcolor=violet]{hyperref}
\usepackage{verbatim}
\usepackage{setspace}
\usepackage{array}
\usepackage{float}
\usepackage{epsfig}
\usepackage{amsbsy}
\usepackage{soul}
\usepackage{xfrac}
\usepackage{enumitem}
\usepackage{multirow}
\usepackage{braket}
\usepackage{mathrsfs}
\usepackage{tikz}

\definecolor{BLUE}{named}{blue}
\usepackage{xspace}
\newcommand{\dosmax}{\texttt{ReSMax}\xspace}

\newcounter{bla}

\journal{Computer Physics Communications}

\begin{document}

\begin{frontmatter}




\title{
An Algorithm for Automated Extraction\\ 
of Resonance Parameters from the Stabilization Method}  


\author[a]{Johanna Langner\textsuperscript{\dag}}
\author[a]{Anjan Sadhukhan}
\author[b]{Jayanta K. Saha}
\author[a]{Henryk A. Witek\corref{hw}}

\address[a]{Department of Applied Chemistry, National Yang Ming Chiao Tung University, 1001 University Road, Hsinchu, 300, Taiwan}
\address[b]{Department of Physics, Aliah University, IIA/27, Newtown, Kolkata 700160, India}

\cortext[hw]{\,Corresponding author. \textit{Email address:} hwitek@nycu.edu.tw}

\date{\today}
\begin{abstract}
The application of the stabilization method [A.~U.\ Hazi and H.~S.\ Taylor, Phys.~Rev.~A {\bf 1}, 1109 (1970)]) to extract accurate energy and lifetimes of resonance states is challenging: The process requires labor-intensive numerical manipulation of a large number of eigenvalues of a parameter-dependent Hamiltonian matrix, followed by a fitting procedure. In this article, we present \dosmax, an efficient algorithm implemented as an open-access \texttt{Python} code, which offers full automation of the stabilization diagram analysis in a user-friendly environment while maintaining high numerical precision of the computed resonance characteristics. As a test case, we use \dosmax to analyze the natural parity  doubly-excited resonance states  (${}^{1}\textnormal{S}^{\textnormal{e}}$, ${}^{3}\textnormal{S}^{\textnormal{e}}$, ${}^{1}\textnormal{P}^{\textnormal{o}}$,  and ${}^{3}\textnormal{P}^{\textnormal{o}}$) of helium, demonstrating the accuracy and efficiency of the developed methodology. The presented algorithm is applicable to a wide range of resonances in atomic, molecular, and nuclear systems.\\  

\end{abstract}
\end{frontmatter}
\let\svthefootnote\thefootnote
\let\thefootnote\relax\footnotetext{
\hangindent=1.8em
\hangafter=1
\noindent\!\!\!\textsuperscript{\dag}\,\textit{For technical questions}: johanna.langner@nycu.edu.tw\\
\textit{For bugs or feature requests:} \href{https://github.com/giogina/ReSMax/issues}{github.com/giogina/ReSMax/issues}}
\addtocounter{footnote}{-1}\let\thefootnote\svthefootnote

\section{Introduction}\label{Introduction}
Atomic and molecular resonances exist within the scattering continuum and have finite lifetimes. In photoabsorption and scattering experiments, these transient metastable states decay via two primary channels: autoionization (emission of an electron, resulting in an ionic configuration) and fluorescence (emission of a photon, resulting in a low-energy excited state) \cite{deHarak2006, saha2013, PhysRevA.98.033416, PhysRevA.110.062802}. Each resonance state is characterized by two main parameters: its energy and lifetime. The lifetime reflects contributions from both the autoionization and fluorescence decay channels, whose relative importance provides a basis for further classification of resonance states \cite{saha2011a}.

Several methods exist for evaluating the resonance parameters of such metastable states, which are generally categorized into two types: direct and indirect methods. Direct techniques include  complex coordinate rotation, complex scaling, and complex absorption potential methods \cite{HO19831, MOISEYEV1998212,PhysRevC.65.054305}, in which the resonance parameters are obtained directly by diagonalizing a complex scaled Hamiltonian. Indirect methods include analytic continuation of energy in the complex plane \cite{1983-mccurdy-CPL, 1981-simos-1981, 1984-isaacson-CPL}, methods using the asymptotic form of the wave function \cite{1970-hazi-PRA, 1988-lefebvre-JPB}, the Feshbach projection operator method \cite{PhysRevA.88.012702}, the R-matrix method \cite{BURKE1973288}, and the stabilization method \cite{1970-hazi-PRA, 1971-Fels-PRA, 1970-Holoien-JPB}.

\begin{figure}[th]
    \centering
    \begin{tikzpicture}
        \node[anchor=south west, inner sep=0] (image) at (0,0) {\includegraphics[width=0.95\linewidth, trim=55 55 0 0, clip]{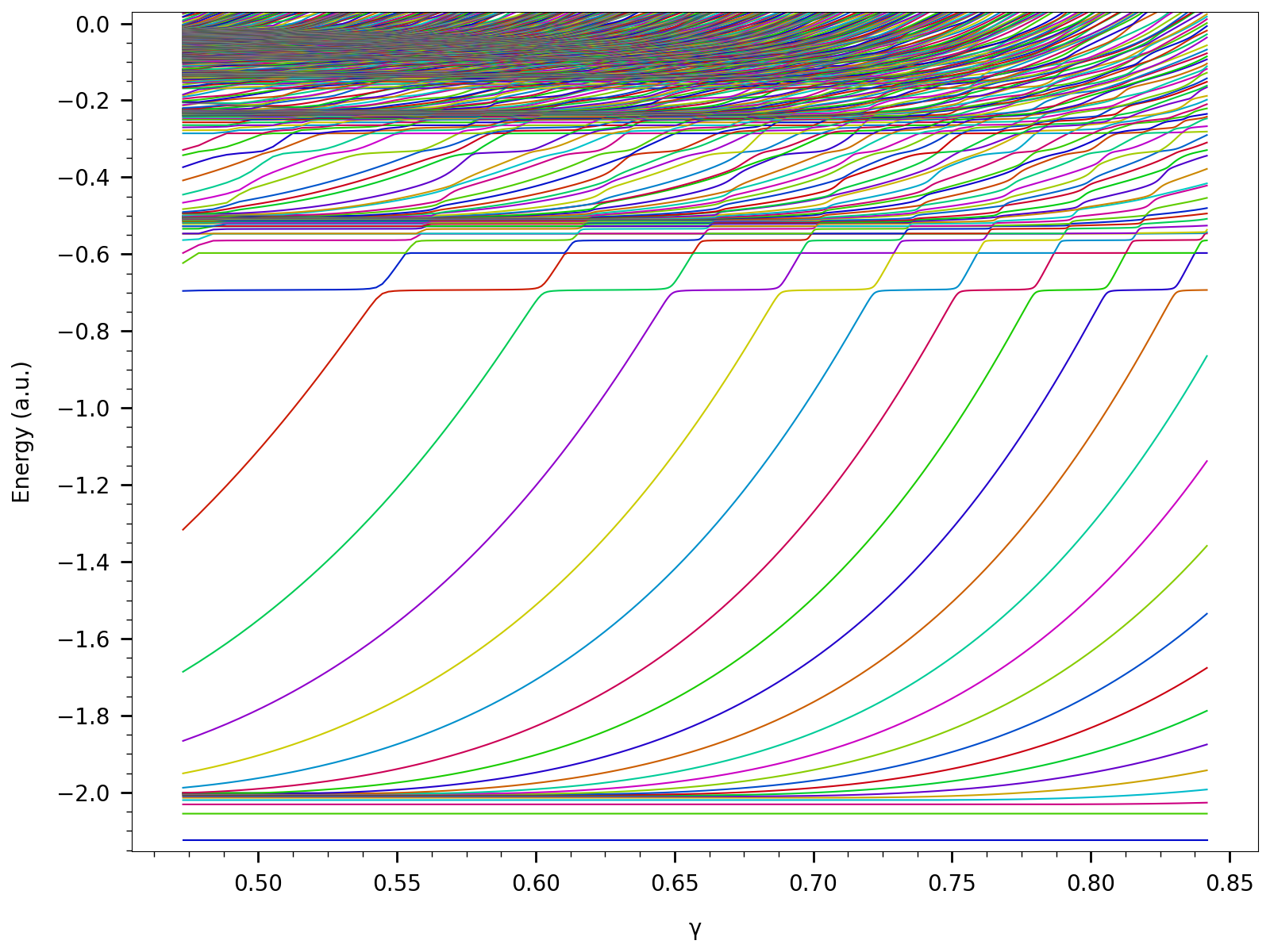}};
        \begin{scope}[x={(image.south east)}, y={(image.north west)}]
            \node[anchor=north] at (0.5, 0) {\scriptsize$\gamma$};
            \node[rotate=90, anchor=south] at (0, 0.5) {\scriptsize$E$ (a.u.)};
        \end{scope}
    \end{tikzpicture}
    \caption{\dosmax-generated stabilization diagram (SD) displaying the first $500$  ${}^{1}\text{P}^{\text{o}}$ energy roots of ${}^{\infty\!}\text{He}$, computed using the basis set given in Eq.~\eqref{rwf} with $n=9$, $\texttt{w}=2$, and $\lambda_{0}=4.0$ a.u., as a function of the basis set parameter $\gamma$. Bound states appear as almost constant lines below $E=-2.0$ a.u.; resonances manifest as plateaus.
    }
    \label{fig:po_he_infm}
\end{figure}

The stabilization method is a widely used indirect approach that avoids the complications associated with complex scaling. It extracts resonance parameters from stabilization diagrams (SDs), which track the dependence of the eigenvalues of the Hamiltonian on a tunable basis set parameter, and can be used to locate bound, resonance and scattering states (see for example Fig.~\ref{fig:po_he_infm}). This parameter probes the completeness of the basis set used in the calculations, and can be defined in different ways. The original implementation of the method \cite{1970-hazi-PRA,1971-Fels-PRA,PhysRevA.12.885} involved varying the size of the basis set. In the so-called hard-wall approach \cite{1980-maier-JPB, 1988-lefebvre-JPB, 1993-Mandelshtam-PRA,PhysRevA.103.012811}, the system is enclosed in a spherical cavity, the radius of which is varied to modify the Dirichlet boundary conditions defining the basis set. A more refined and stable variant, the soft-wall approach \cite{1994-Muller-PRA,kar2005,saha2009}, was later introduced, in which a specific parameter within the basis set of a fixed size is systematically varied to probe the dependence of energy levels on the radial extent of the corresponding wave functions. The soft-wall strategy has proven highly effective in predicting resonance parameters across various systems, including free, confined, and field-induced few-body systems, for which Hylleraas-type basis sets are typically used \cite{kar2004,kar2005,kar2005a,kar2006,kar2007,ghoshal2009,saha2009,saha2010, saha2011, saha2013, Saha_2016, Sadhukhan2019, dutta2019}. The method has also been extended to determine atomic resonances near metal surfaces \cite{nimb1995}, resonances in molecules \cite{landau2019,GonzalezLezana2002}, core-excited shape resonances in clusters \cite{fennimore2016,Fennimore2018}, as well as to probe nuclear resonances \cite{Zhang2008}. 

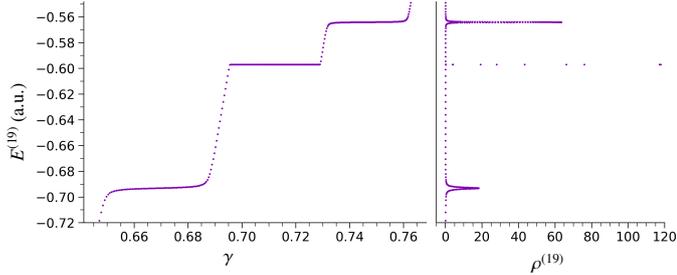
\begin{figure}[bh]
    \centering
    \hspace*{-0.03\linewidth}
    \noindent\makebox[1.03\linewidth][c]{
        \begin{tikzpicture}
            \node[anchor=south west, inner sep=0] (image) at (0,0) {\includegraphics[width=0.95\linewidth]{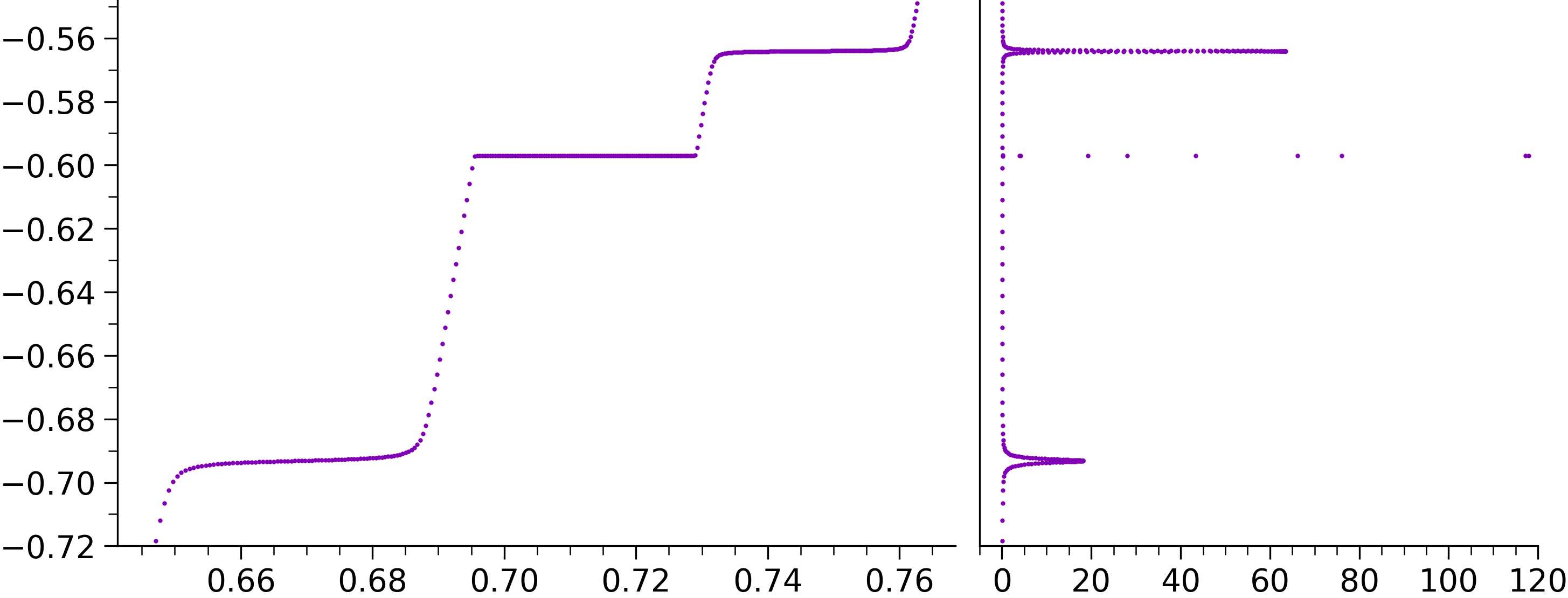}};
            \begin{scope}[x={(image.south east)}, y={(image.north west)}]
                \node[rotate=90, anchor=south] at (0.00, 0.5) {\scriptsize$E^{(19)}$\,(a.u.)};
                \node[anchor=north] at (0.3, 0) {\scriptsize$\gamma$};
                \node[anchor=north] at (0.8, 0) {\scriptsize$\rho^{(19)}$};
            \end{scope}
        \end{tikzpicture}
    }
    \caption{Energy curve $(\gamma_i, E^{(19)}_i)$ (left) and density of states (DOS) curve $(E^{(19)}_i, \rho_i^{(19)})$ (right, rotated by 90$^{\circ}$ to share the energy axis) for \textcolor[HTML]{8000B2}{root 19} of ${}^{1}\text{P}^{\text{o}}$  ${}^{\infty\!}\text{He}$ (compare Fig.~\ref{fig:po_he_infm}) over the energy range from $-0.72$ to $-0.55\text{ a.u.}$ 
    The energy curve exhibits three plateaus at $E_r = -0.693$, $-0.597$ and $-0.564$ a.u. Each plateau indicates a resonance state, and corresponds to a peak in the DOS curve centered at the same energy. Flatter plateaus produce narrower and higher DOS peaks, indicating longer lifetimes.
    (The DOS peak at $-0.597$ exceeds the plotted $\rho^{(19)}$ range by a factor of 47.)
    }
    \label{fig:plateaus-to-dos-peaks}
\end{figure}

The stabilization method is capable of distinguishing between bound, resonance, and scattering solutions by analyzing the characteristic changes of the energy eigenvalues observed during systematic variation of the designated basis set parameter.  Bound state energies remain relatively unaffected in this process, while scattering states vary monotonically with the parameter, and resonance states (RSs) show both monotonic variation and stabilization within specific parameter ranges. Each energy value at which the roots form such plateaus is taken as the resonance position $E_r$ of a corresponding RS.
Mandelshtam \textit{et al.}~\cite{1993-Mandelshtam-PRA} and M\"{u}ller \textit{et al.}~\cite{1994-Muller-PRA} put forward an elegant and simple method to determine the resonance parameters (position and lifetime) from an SD.  The procedure begins with the determination of the density of states $\rho^{(n)}(E)$ associated with each root $n$. In practical implementations, where the basis set parameter $\gamma$ runs over a finely spaced array of $K$ consecutive values, $\texttt{gamma} = \left[\gamma_0,\ldots,\gamma_K \right]$, giving rise to an analogous array  $\texttt{energy[$n$]} = [E^{(n)}_0,\ldots,E^{(n)}_K ]$ of eigenvalues corresponding to the root $n$, the corresponding density of states (DOS) can be estimated by the array $\texttt{rho[$n$]} = [\rho^{(n)}_1,\ldots,\rho^{(n)}_{K-1}]$ with the values $\rho^{(n)}_i$ computed as 
\begin{equation}\label{eq:DOS}
\rho^{(n)}_i = \frac{\gamma_{i+1}-\gamma_{i-1}}{E^{(n)}_{i+1}-E^{(n)}_{i-1}}
\end{equation}
The DOS curves $(E^{(n)}_i, \rho_i^{(n)})$ exhibit readily identifiable peaks at energies where the $(\gamma_i, E^{(n)}_i)$ curves form plateaus, as shown in Fig.~\ref{fig:plateaus-to-dos-peaks}. Each of these peaks can be locally fitted to a Lorentzian  of the form
\begin{equation}\label{eq:lorentzian}
L(E)=y_0+\frac{A}{\pi} \frac{\Gamma/2}{(E-E_r)^2+(\Gamma/2)^2}
\end{equation}
where $y_0$ is the baseline offset and $A$ denotes the total area under the curve. The two most important quantities extracted from this fit are the position of the resonance $E_r$ and its width $\Gamma$, which is inversely proportional to its lifetime $\tau$. In situations when the same resonance state manifests itself as a plateau for multiple roots, the resonance parameters $E_r$ and $\Gamma$ are selected from the best-fitting DOS peak.

Despite its potential and its conceptual simplicity, the stabilization method remains challenging to apply due to one main difficulty. The manual workflow for resonance extraction is a tedious and error-prone task: Calculating the DOS for each root, fitting each DOS peak using standard data visualization software, and finally selecting the best fit for each resonance is time-consuming and susceptible to human error, especially when a large basis sets are considered. In practice, it can take several minutes or even an hour of human time to analyze each RS, depending on the complexity of the SD, making large-scale studies time-consuming and difficult to reproduce consistently.

To overcome these limitations, we introduce \dosmax, an automated command-line tool designed to streamline the detection of bound and resonance states with minimal human intervention. Given an SD as input, \dosmax\ identifies and fits DOS peaks, assigns them to resonances, and determines the resonance parameters --- all within a matter of seconds. The tool offers interactive options to adjust peak selections based on real-time generated plots. \dosmax greatly reduces computation time, eliminates subjective bias, and ensures consistent results. The software is open source and available, along with documentation and installation instructions, at \cite{Langner2025}.

To showcase both the accuracy and the automation efficiency of \dosmax, we apply it to the doubly excited RSs of helium, a well-studied Coulomb few-body system with rich resonance structure. Due to the availability of explicitly correlated techniques, Coulomb few-body systems provide an ideal benchmark for testing the proposed algorithm. We test \dosmax on the natural parity  doubly-excited ${}^{1}\textnormal{S}^{\textnormal{e}}$, ${}^{3}\textnormal{S}^{\textnormal{e}}$, ${}^{1}\textnormal{P}^{\textnormal{o}}$,  and ${}^{3}\textnormal{P}^{\textnormal{o}}$ resonance states of the helium atom (for both infinite and finite nuclear mass) below the He$^+$($2s$) threshold. The results are compared with existing high-precision theoretical estimates and the limited experimental data available.

In this paper, we first describe the evaluation of the energy eigenroots as a function of the basis-set parameter, using the explicitly correlated Ritz variational framework in Section~\ref{method}.  In Section~\ref{algorithm}, we explain the algorithm used by \dosmax to systematically analyze an SD to extract resonance parameters. Section~\ref{dataextraction} demonstrates the workflow of \dosmax through a practical example. Finally, the resulting resonance parameters for the  ${}^{1}\textnormal{S}^{\textnormal{e}}$, ${}^{3}\textnormal{S}^{\textnormal{e}}$, ${}^{1}\textnormal{P}^{\textnormal{o}}$,  and ${}^{3}\textnormal{P}^{\textnormal{o}}$ states of the helium atom are presented in Section~\ref{results}.
 
\section{Computation of Energy Eigenroots}\label{method}

To detect doubly-excited RSs of the helium atom, we solve the time-independent Schr\"{o}dinger equation (SE) for the system. For a given total orbital angular momentum $L$ and its projection $M$ onto the laboratory-fixed $z$-axis, the SE of the helium atom takes the following form in the nucleus-centered coordinate frame.
\begin{equation}\label{secm0}
        \left[-\frac{1}{2\mu}\sum_{i=1}^{2}\Delta_{i}-\frac{1}{m}\nabla_{1}\cdot\nabla_{2}+\mathcal{V}-E \right]\,\Psi^{L}_{\!M}(\bm{r}_{1},\bm{r}_{2}) = 0
    \end{equation}
Here, $E$ refers to the internal energy of the system, $m$ is the mass of the nucleus, and $\mu=\frac{m}{1+m}$ is the reduced mass of the two quasi-particles, located at positions $\bm{r}_{1}$ and $\bm{r}_{2}$ relative to the nucleus. The potential energy operator $\mathcal{V}$ is given by
\begin{equation}
    \mathcal{V} = -\frac{2}{r_{1}}-\frac{2}{r_{2}} +\frac{1}{r_{12}}
\end{equation}
where $r_1=\left| \bm{r}_{1} \right|$, $r_2=\left| \bm{r}_{2} \right|$, and $r_{12}=\left| \bm{r}_{1} - \bm{r}_{2} \right|$ are the inter-particle distances. The Hamiltonian in Eq.~\eqref{secm0} includes the mass-polarization term $-\frac{1}{m}\nabla_{1}\cdot\nabla_{2}$, which vanishes in the limit of infinite nuclear mass. For natural parity, given by $(-1)^L$, the wave function $\Psi^{L}_{M}(\bm{r}_{1},\bm{r}_{2})$ is expanded in terms of the parity-adapted solid minimal bipolar harmonics (MBH) $\bm{\Omega}_{l}^{LM}(\bm{r}_{1},\bm{r}_{2})$ as \cite{Sadhukhan2025}
\begin{equation}\label{wfa1}
    \Psi^{L}_{\!M}(\bm{r}_{1},\bm{r}_{2}) = \sum_{l=0}^{L}\psi_{l}^{L}(r_{1},r_{2},r_{12})\,\bm{\Omega}_{l}^{LM}(\bm{r}_{1},\bm{r}_{2}) \pm \text{exchange}
\end{equation}
where the partial wave components $\psi_{l}^{L}(r_{1},r_{2},r_{12})$ are functions of the inter-particle distances $r_{1}$, $r_{2}$ and $r_{12}$. The exchange term arises from the permutation of the two quasi-particles, ensuring the proper symmetry of the wave function. The $\pm$ sign corresponds to spin-singlet and spin-triplet states, respectively. In bispherical coordinates $(r_{1},\theta_{1},\phi_{1},r_{2},\theta_{2},\phi_{2})$, solid MBH take the form \cite{Sadhukhan2025}
\begin{equation}\label{bsheqn0}
\bm{\Omega}_{l}^{LM}(\bm{r}_{1},\bm{r}_{2})\hspace{-2pt} =\hspace{-2pt} r_{1}^l r_{2}^{L-l}\hspace{-3pt}\sum_{m=\kappa_{1}}^{\kappa_{2}}\hspace{-2pt}\mathcal{C}_{l,m,L-l,M-m}^{LM}Y_{m}^{l}(\theta_{1},\phi_{1})~ Y_{M-m}^{L-l}(\theta_{2},\phi_{2})
\end{equation}
where $\kappa_{1}=\text{max}(-l,l-L+M)$, $\kappa_{2}=\text{min}(l,L+M-l)$, $-L\leq M \leq L$, $0\leq l \leq L$, $\mathcal{C}_{l,m,L-l,M-m}^{LM}$ are the Clebsch-Gordan coefficients, and $Y_{m}^{l}(\theta,\phi)$ are the spherical harmonics.

In order to apply the soft-wall strategy, we need to expand the wave function in a basis set that depends on a given parameter (say $\gamma$) which can be systematically varied to generate SDs. This is achieved by expanding the partial wave components $\psi_{l}^{L}\left(r_{1}, r_{2}, r_{12}\right)$ in an explicitly correlated multi-exponent Hylleraas-type basis set  \cite{Hylleraas1928, Hylleraas1929} as
\begin{equation}\label{rwf}
    \psi_{l}^{L}\left(r_{1}, r_{2}, r_{12}\right) = \sum_{i=1}^{\texttt{N}_a}\sum_{j=1}^{\texttt{N}_b} \mathscr{C}_{lij}^{L}~{r_{1}^{\texttt{l}_{j}} r_{2}^{\texttt{m}_{j}} r_{12}^{\texttt{n}_{j}}}{e^{-\alpha_{i} r_{1}-\beta_{i} r_{2}}}
\end{equation}
where the index $i$ enumerates $\texttt{N}_a$ sets of positive non-linear variational coefficients ($\alpha_{i}, \beta_{i}$), $j$ enumerates $\texttt{N}_b$ sets of non-negative integers ($\texttt{l}_{j}$, $\texttt{m}_{j}$, $\texttt{n}_{j}$), and $\mathscr{C}_{lij}^{L}$ are the linear variational coefficients to be solved for. The exponent pairs ($\alpha_{i}, \beta_{i}$) are constructed as the two-element subsets of the geometric progression (GP),
\begin{equation}\label{gpr}
    \{\alpha_i, \beta_i\} \subset \{ \lambda_0 \gamma^i ~\vert~ i=0,\ldots,n-1 \},
\end{equation}
resulting in 
\begin{equation}\label{Na}
    \texttt{N}_a=\binom{\,n\,}{2}
\end{equation}
unique pairs of exponents. Here, $\lambda_0$ is the first term of the GP series \eqref{gpr} that can be selected according to the size of the given system (for the present problem we chose $\lambda_0 = 4$). The GP ratio $\gamma$ acts as the basis set parameter for the soft-wall method, since its smooth variation, typically within $0.4 < \gamma < 0.9$, can induce stabilization of eigenroots. The distinct sets of positive integers ($\texttt{l}_{j}$,$\texttt{m}_{j}$,$\texttt{n}_{j}$) are determined by applying the constraint $\texttt{l}_{j}+\texttt{m}_{j}+\texttt{n}_{j}\leq \texttt{w}$ for a given integer parameter \texttt{w}. Therefore, the number $\texttt{N}_b$ of such sets is
\begin{equation}\label{Nb}
    \texttt{N}_b=\binom{\,\texttt{w}+3\,}{3}
\end{equation} 
For a given value of $L$, the total number of terms in the wave function \eqref{wfa1} is given by
\begin{equation}\label{Nprod}
    \texttt{N}=\texttt{N}_a \,\texttt{N}_b \,(L+1)
\end{equation}

The Rayleigh-Ritz variational principle is employed to obtain the energy eigenvalues for a given $\gamma$ value. This reduces the SE in Eq.~\eqref{secm0} to a generalized eigenvalue problem (GEP) of the form
\begin{equation}\label{gep}
\mathbb{H}\, c = E\, \mathbb{S}\, c
\end{equation}
In this formulation, $\mathbb{H}$ is the Hamiltonian matrix, $\mathbb{S}$ is the overlap matrix, and $c$ is the column vector of the linear expansion coefficients of the wave function~\eqref{rwf}. The matrix elements are evaluated analytically over the basis functions using the closed-form integral expressions reported by Calais and L\"{o}wdin \cite{Calais1962}. The overall methodology, including the construction of these matrices and the variational framework, follows the approach described in details in \cite{Sadhukhan2025}. All numerical calculations described in this section are performed using a robust, quadruple precision  code \cite{Sadhukhan2025} written in  \texttt{Julia}  \cite{julia_bezanson_2017}.

\section{Data Analysis Algorithm}\label{algorithm}
The core algorithm used by \dosmax to analyze SDs performs the following operations. It parses the input eigenroot spectrum (Subsection~\ref{subsec:input-file-parsing}), computes the DOS (Subsection~\ref{subsec:dos-calc}), identifies plateau regions corresponding to resonances and fits the associated DOS peaks to Lorentzian curves (Subsection~\ref{subsec:peak-fitting}), selects the best-fitting peak for each resonance (Subsection~\ref{subsec:resonance-identification}), and outputs the resonance parameters in a structured format (Subsection~\ref{subsec:adjustments-output}).
While the analysis is fully automated, \dosmax\ also provides an interactive console interface for reviewing and refining peak selections. Users may replot data, modify energy ranges, and refine DOS peak selection. The results of the analysis are provided in a tabular text file listing the extracted resonance parameters. Additionally, various plots can be generated and saved as \texttt{.png} images.

The software is implemented in \texttt{Python 3.10} \cite{python}, released under the MIT license, and available open source at \href{https://github.com/giogina/ReSMax}{\texttt{github.com/giogina/ReSMax}}, which also provides installation instructions and a user guide. A detailed demonstration of its application is given in Subsection \ref{dataextraction}.

\subsection{Input File Parsing}\label{subsec:input-file-parsing}
The input file with the eigenroot spectrum across a range of the tunable basis set parameter $\gamma$, in our case automatically generated by the   \texttt{Julia}  code \cite{Sadhukhan2025}, can be provided in one of three supported formats:

\begin{itemize}
    \item \textbf{Tabular format (.dat):} A tab-separated file where each row contains a $\gamma$ value followed by energy values for multiple roots.
    \item \textbf{Block-structured format (.dal):} A structured format where each block begins with a $\gamma$ value, followed by lines listing energy values for each root, with blocks separated by blank lines.
    \item \textbf{Array-based format (.ou):} A format where the first line lists the $\gamma$ values, and the following lines list the consecutive energy values for all the roots.
\end{itemize}

The format is detected automatically based on the file extension. If an unrecognized format is encountered, the program issues a warning and exits.  All output files --- including the plots described below as well as the \texttt{results.txt} file summarizing the detected bound and resonance states --- are stored in a directory named after the input file, created in its parent directory.

The parsed data is stored in a standardized dictionary called \texttt{data}, with the array of $\gamma$ values under the key \texttt{"gamma"} and the array of energies of each root stored under an integer key corresponding to the root number. This data is then visualized as an SD using \texttt{Matplotlib} \cite{Matplotlib2007}, as shown in Fig.~\ref{fig:po_he_infm}.

After parsing and before initiating automatic detection of bound and resonance states, the user may input manually the hydrogenic ionization thresholds (HITs), which otherwise are set to the default sequence $-Z^2/(2n^2)$ with $n = 1, \ldots, 20$ and $Z=2$.

\subsection{DOS Calculation}\label{subsec:dos-calc}
The DOS calculation for root \texttt{n} according to Eq.\,\eqref{eq:DOS} is implemented efficiently as a single \texttt{NumPy} \cite{harris2020array} array operation:
\begin{equation} \label{dos_numpy}
    \texttt{rho[n]} = \frac{\texttt{gamma[2:]} - \texttt{gamma[:-2]}}{\texttt{energy[n][2:]} - \texttt{energy[n][:-2]}}
\end{equation}
Here, \texttt{gamma=data["gamma"]} and \texttt{energy[n]=data[n]} are the \texttt{NumPy} arrays of $\gamma$ and energy values retrieved from the \texttt{data} dictionary. The slicing notation used in this equation functions as follows:

\(\texttt{gamma[2:]}=[\gamma_2, \gamma_3, \ldots, \gamma_i]\) skips the first two entries,

\(\texttt{gamma[:-2]}=[\gamma_0, \ldots, \gamma_{i-2}]\) skips the last two entries.

\noindent Equivalent slicing is applied to the energy array. \texttt{NumPy} applies Eq.\,\eqref{dos_numpy} element-wise, yielding an array of DOS values $\rho$. The resulting array is stored in \texttt{data["rho\_n"]}.

\subsection{Detection of Bound States}\label{subsec:BS-detection}
Bound states are detected in a two-step process. 1)~The global energy minimum $E_{\min}$ for each root $n$ is located. 2)~If this minimum lies below the lowest HIT, it is recognized as a bound state and listed together with the corresponding $\gamma$ value. A warning is issued if $E_{\min}$ occurs within the first or last five points of the array, since this suggests that the true minimum may lie outside the sampled $\gamma$ range.

\subsection{Peak Identification and Fitting}\label{subsec:peak-fitting}
To systematically identify peaks in the DOS, \dosmax pre-processes the data as follows:
\begin{description}[style=unboxed,leftmargin=0cm]
    \item [Data segmentation] To isolate individual peaks, the data arrays are segmented at the points of steepest energy increase w.r.t.\ $\gamma$ (i.e., rising points of inflection in \texttt{energy[n]}), which correspond to local minima of the DOS satisfying $\rho > 0$. For each segment, the energy subarray together with the corresponding $\gamma$ and $\rho$ subarrays are assigned to an instance of the \texttt{DOSpeak} class, which handles subsequent processing.
    
    \item [Trimming discontinuities] To prevent outliers or abrupt changes in DOS from distorting the subsequent Lorentzian fit, each segment is trimmed to retain only the continuous portion of the peak. Specifically, the segment with a maximum DOS value of $\rho_{\text{m}}$ is considered discontinuous at $i$ if the change in DOS satisfies $\vert\rho_{i+1}-\rho_i\vert>0.6\, \rho_{\text{m}}$. Then, the segment is split after the index $i$, and only the subsegment that includes $\rho_{\text{m}}$ is retained.
    
    \item [Discarding bad segments] Segments are discarded entirely if they are unlikely to yield a reliable fit. Specifically, a segments is removed if: 1)~it contains fewer than 10 data points, 2)~its maximum DOS value occurs at the first or last entry of the $\rho$ array, or 3)~it exhibits significant asymmetry between the steepest ascent and descent of the DOS curve.

    \item [Estimating Lorentzian fit parameters]  The peak position $E_{\text{m}}$ and height $\rho_{\text{m}}$ are approximated by the energy and DOS values at the maximum of the $\rho$ array. The width $\Gamma_{\text{m}}$ is estimated from the separation between half-maximum points, while the corresponding area $A_{\text{m}}$ is derived from Eq.~(\ref{eq:lorentzian}). The baseline $y_{0,\text{m}}$ is initially set to half of the minimum DOS value and then adjusted to better match the DOS at the midpoint of the segment.

    \item [Zooming in on the peak] Based on the initial parameter estimates, the data arrays are trimmed to retain only the region $E_{\text{m}} \pm 10\, \Gamma_{\text{m}}$. This focuses the fitting process on the most relevant portion of the peak, improving accuracy and efficiency.

\end{description}

After identifying and trimming, each DOS peak segment is fitted to a standard Lorentzian curve given by Eq.~\eqref{eq:lorentzian}, using \texttt{scipy}'s \texttt{curve\_fit} routine \cite{2020SciPy-NMeth} with the Trust Region Reflective algorithm. The fitting process, based on non-linear least squares optimization, determines the parameters \(y_0\), \(A\), \(\Gamma\), and \(E_r\) by minimizing the sum of squared residuals (SSR)
\begin{equation}\label{eq:ssr}
    \text{SSR} = \sum_{i=1}^{N} \left( \rho(E_i) - L(E_i) \right)^2
\end{equation}
where \(\rho(E_i)\) represents the DOS data points and \(L(E_i)\) the Lorentzian function values at those points. The fitting allows up to three attempts with adjusted starting parameters. The fit is retained only if the fitted peak position lies within an acceptable range of the segment maximum $\rho_{\text{m}}$. In cases when no acceptable fit can be obtained in this way, the analyzed segment is discarded from further analysis.

In addition to the SSR and the optimized parameters \(y_0\), \(A\), \(\Gamma\), and \(E_r\), the relative SSR per point defined as
\begin{equation}\label{eq:rsspp}
    \chi_{r} = \frac{\sqrt{SSR}}{N \cdot \rho_{\text{max}}^2}
\end{equation}
is stored in the \texttt{DOSpeak} class. The $\chi_{r}$ value allows for fair comparison of fits across DOS peaks with varying heights and data point counts, providing a more robust measure than the simple SSR value. Dividing by the square of the peak height prioritizes higher DOS peaks, which are more likely to represent a RS well.

\begin{figure}[th]
    \centering
    \includegraphics[width=1.00\linewidth]{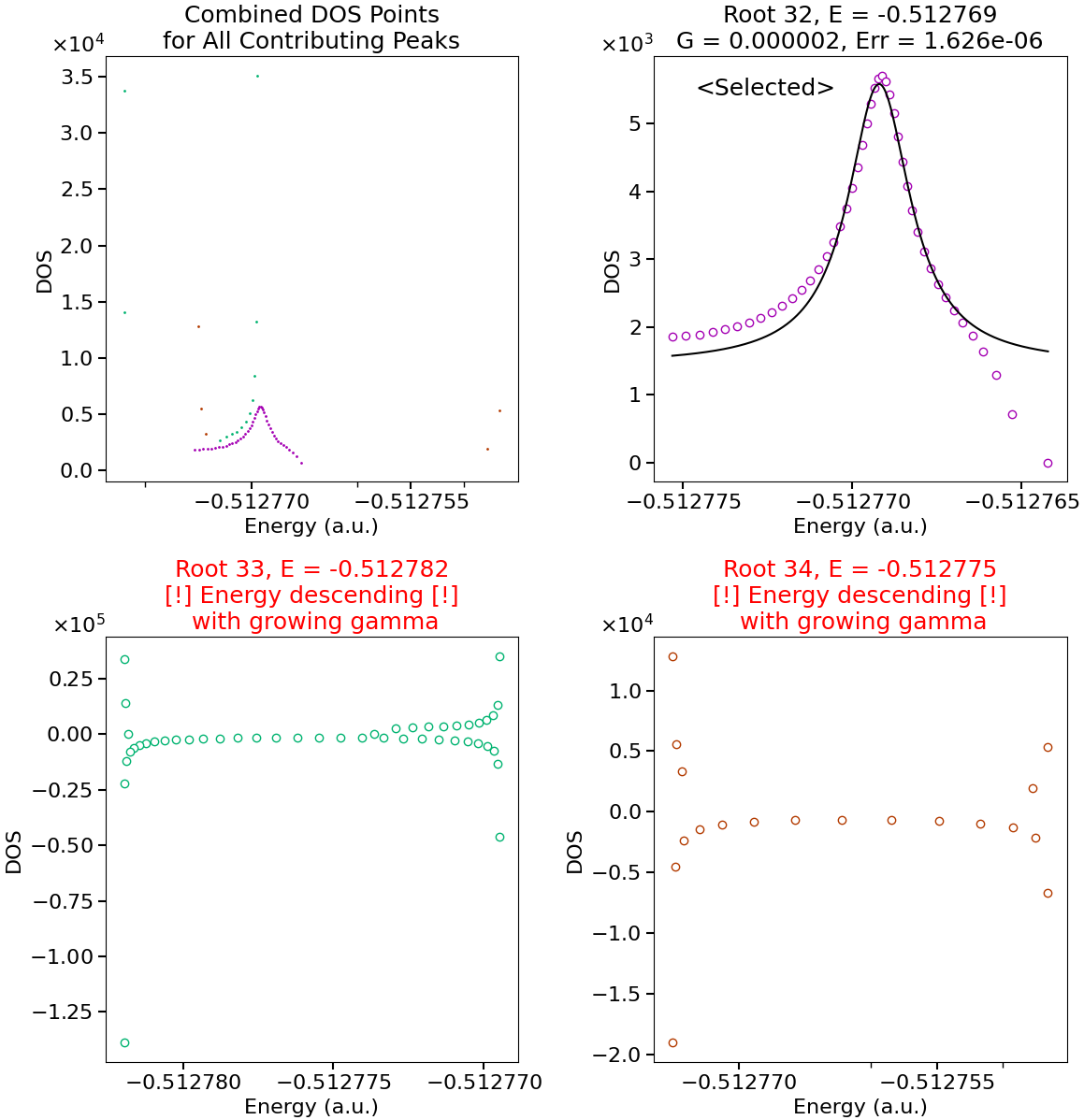}
    \caption{Grid plot displaying the DOS points and corresponding Lorentzian fits (black line) of roots \textcolor[HTML]{AB07B9}{32}, \textcolor[HTML]{07B9B3}{33}, and \textcolor[HTML]{B96A07}{34} for the $13^\text{th}$ RS of ${}^1\text{P}^{\textnormal{o}}$ ${}^{\infty}\text{He}$. Only for root \textcolor[HTML]{AB07B9}{32}, a Lorentzian fit is viable; and the Lorentzian curve does not match the DOS data points particularly well. This shows that RS 13 can not be reliably detected using the basis set at hand.}
    \label{fig:bad_fit_grid}
\end{figure}

\begin{description}[style=unboxed,leftmargin=0cm]
    \item [Descending energy plateaus] In some cases, the energy does not increase monotonically within an energy plateau in the SD. There are two main reasons for this phenomenon: 1)~An insufficient size of the basis set used to model an RS with a high angular momentum, which can result in inadequate stabilization patterns of the energy eigenroots. In the case of doubly excited RS of helium, this effect can be observed directly below the HITs, as seen e.g.\ in Fig.~\ref{fig:resonance_overview_not_nice}. 2)~For fluorescence-active resonances with long autoionization lifetimes (represented in the SD by nearly constant energy plateaus), small numerical artifacts arising from incompleteness of the basis set can result in a subtle decrease in energy. Since the DOS computed with Eq.~(\ref{eq:DOS}) and associated with descending energy plateaus does not form a peak (compare Fig.~\ref{fig:bad_fit_grid}), the fitting procedure described above can not be performed. Instead, \dosmax assigns to such segments a position corresponding to their minimum energy $E_\text{min}$ and a large fitting error of $\chi_r = 10^6 - E_\text{min}$, which deprioritizes them during resonance selection.
\end{description}

\subsection{Identification of Resonances}\label{subsec:resonance-identification}
The previous operation produces a list of instances of the \texttt{DOSpeak} class, each representing either a DOS peak with a valid Lorentzian fit, or a descending energy plateau marked by a large fitting error. In the following, both types are treated equivalently and referred to simply as “peaks”.
These peaks are now grouped into resonances, with the best-fitted peak in each group selected as its representative, as follows:
\begin{description}[style=unboxed,leftmargin=0cm]
    \item [Threshold categorization] The energy ranges between successive HITs are treated as distinct regions for identifying resonances. In this way, each peak is assigned to an unique energy interval between two successive HITs.

    \item [Peak grouping] Within each such energy interval, the program iterates over all peaks in ascending order of peak energy $E_r$. Peaks are added to a temporary buffer until two peaks in the buffer originate from the same root.
    When this occurs, the largest energy gap between consecutive peaks in the buffer is identified. All peaks up to this gap are assigned to a new instance of the \texttt{Resonance} class and removed from the buffer. The remaining peaks (if any) stay in the buffer, and the process continues.
\end{description}

Each such new instance of the \texttt{Resonance} class corresponds to one RS.  The peak with the lowest $\chi_r$ is designated as the resonance’s representative, and the resonance parameters are taken from this representative.
Resonances with fewer than 3 contributing peaks are discarded.

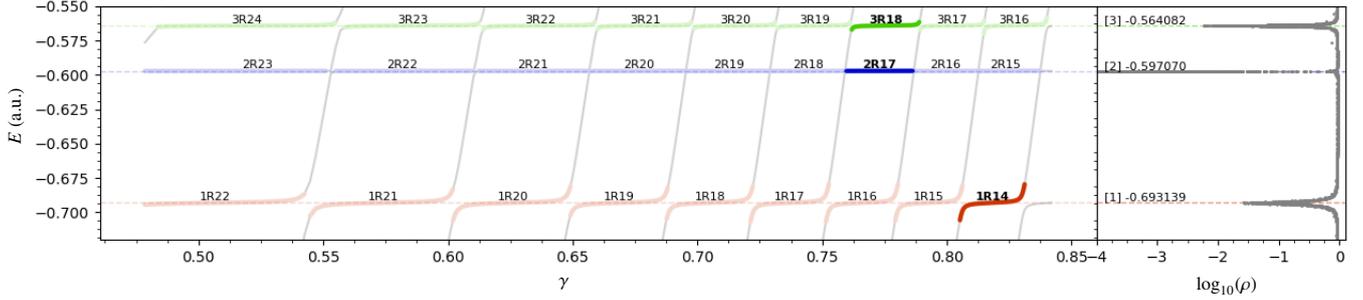
\begin{figure*}[th]
    \centering
    \begin{tikzpicture}
        \node[anchor=south west, inner sep=0] (image) at (0,0) {\includegraphics[width=0.95\linewidth, trim=30 20 0 0, clip]{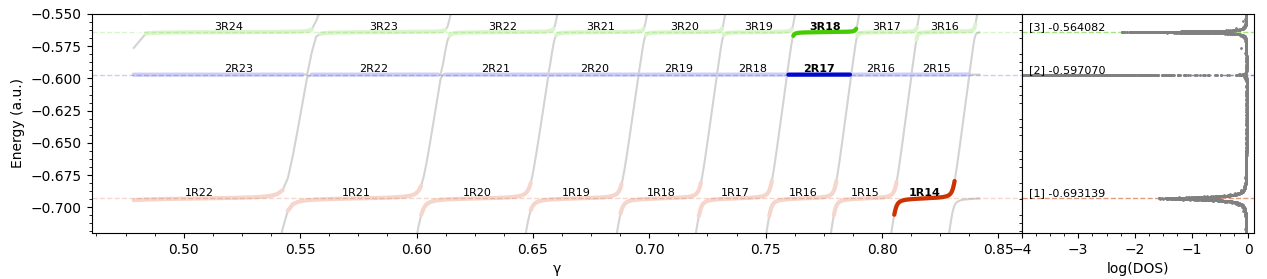}};
        \begin{scope}[x={(image.south east)}, y={(image.north west)}]
            \node[anchor=north] at (0.4, 0) {\scriptsize$\gamma$};
            \node[anchor=north] at (0.9, 0) {\scriptsize$\log_{10}(\rho)$};
            \node[rotate=90, anchor=south] at (0, 0.55) {\scriptsize$E$ (a.u.)};
        \end{scope}
    \end{tikzpicture}
    \caption{\dosmax-generated resonance overview plot corresponding to the three lowest ${}^{1}\text{P}^{\text{o}}$ resonance states of ${}^{\infty\!}\text{He}$, consisting of a stabilization diagram on the left with highlighted plateau regions and detected resonance states (RS), and a scatter plot on the right with the DOS distribution for each roots, confirming the resonance positions.}
    \label{fig:sd_dos_he_infm_enlarged}
\end{figure*}

\subsection{User Adjustments and Reporting}\label{subsec:adjustments-output}
After automatic resonance detection, the program enters an interactive input loop, described in detail in Section~\ref{dataextraction}. This allows users to review, validate, and, if necessary, adjust the representative peak for each resonance. After this verification step, the final parameters of all accepted resonances are written to the tab-separated output file \texttt{results.txt}. This includes: the resonance position $E_r$, width $\Gamma$, peak amplitude $A$, and baseline offset $y_0$, as well as the basis set parameter $\gamma$ at which the DOS peak occurs. Additionally, plots of all representative peaks, together with their respective Lorentzian fits can be batch-generated (e.g.~Fig.~\ref{fig:fit_1}).

\section{Example application of \protect{\dosmax}}\label{dataextraction}
This section demonstrates how to use \dosmax to analyze the $^{1}\text{P}^\text{o}$ states of the helium atom with infinite nuclear mass, using data generated as described in Section~\ref{method}. The basis set is constructed according to Eq.~\eqref{rwf} with $n=9$, $\texttt{w}=2$, and $\lambda_{0}=4.0$, resulting in a basis set size of $\texttt{N}=720$. For each of $1000$ values of the tunable basis set parameter $\gamma$ (ranging from $0.4729$ to $0.8458$), the lowest $500$ energy eigenvalues are listed in the input file \texttt{he\_1Po\_InfMass.dal}, available in the directory \texttt{example/}  on GitHub \cite{Langner2025}.
This file is processed using the command
\begin{verbatim}
python resmax.py -f /path/to/he_1Po_InfMass.dal
\end{verbatim}
The process outlined in this section reproduces the first two columns of Table~\ref{tab:rsp_p-state}. An extensive user guide on how to use \dosmax is given in the \texttt{README} file on \texttt{GitHub} \cite{Langner2025}.

\subsection{Automatic Resonance and Bound State Detection}\label{demo-Sd}
Upon parsing the input (Subsection~\ref{subsec:input-file-parsing}), \dosmax generates the SD shown in Fig.~\ref{fig:po_he_infm}.
In the present example, the default HITs ($-Z^2/n^2/2\text{ a.u.}$ with $Z=2$ and $n=1,\ldots,20$) are used. 
Once the automatic analysis is initiated by entering the command \texttt{r}, \dosmax executes the steps described in the previous section:
It computes the DOS for each root (Subsection~\ref{subsec:dos-calc}), then detects bound states by identifying global energy minima of the nearly constant roots observed below the lowest ($n=1$) HIT at $-2.0\text{ a.u.}$ (Subsection~\ref{subsec:BS-detection}). Then, \dosmax identifies the resonance states indicated by the plateaus in the SD by fitting Lorentzian curves to the associated DOS peaks,  grouping the peaks into resonances and selecting a representative peak for each of them (Subsections~\ref{subsec:peak-fitting}--\ref{subsec:resonance-identification}). 
For the currently considered example dataset, input file parsing takes approximately 1 second, and automatic resonance detection completes in about 2.5 seconds on a Lenovo T470 laptop (Intel i5-6300U CPU, 8GB RAM).

\subsection{Inspection of Detected Resonances}

While \dosmax automatically detects and categorizes resonances, certain situations --- such as cases of closely spaced plateaus or regions where resonances morph into scattering states as $\gamma$ increases --- may require user intervention. To address such situations, after automatic detection, the program switches to interactive mode, allowing users to inspect and manually adjust the resonances.

For each HIT, the user is presented with a \textit{resonance overview plot} consisting of two subplots: an SD on the left, where identified plateau regions and resonances are highlighted, and a DOS clustering scatter plot on the right. Fig.~\ref{fig:sd_dos_he_infm_enlarged} shows a section of this plot for the energy interval  $[-0.72~\text{a.u.},-0.55~\text{a.u.}]$, illustrating three well-behaved RSs. Each resonance is marked in a distinct color, and each of its plateaus is annotated with a symbol \texttt{iRj}, where \texttt{i} signifies the resonance number and \texttt{j} is the root number. The plateau corresponding to the representative DOS peak of each resonance is marked in a darker shade and annotated in bold. For instance, in Fig.~\ref{fig:sd_dos_he_infm_enlarged}, eigenroot number \texttt{j}=17 is selected as the representative of the second RS ($\texttt{i}=2$). The resonance energy $E_r$, derived from the representative DOS peak, is indicated using a horizontal dashed line and annotated on the right. The displayed resonances are the ${}^{1}\text{P}^{\text{o}}$ states 1, 2 and 3 of ${}^{\infty\!}\text{He}$ listed in Table~\ref{tab:rsp_p-state} with resonance energies $E_r = -0.69314$, $E_r = -0.59707$, and $E_r = -0.56408$ a.u. The associated clustering scatter plot on the right side of Fig.~\ref{fig:sd_dos_he_infm_enlarged} shares the SD's vertical energy axis and displays $\log_{10}(\rho)$ along the horizontal axis. The resulting scatter plots allow us to recognize resonances even in densely populated energy regions, as seen in Fig.~\ref{fig:resonance_overview_not_nice}. Conversely, regions where the scatter plots show random noise indicate that no reliable resonances can be detected there, as illustrated in the next subsection.

The resonance overview plot can be adjusted to display any desired energy range. An optional panoramic version of the $\log_{10}(\rho)$ scatter plot covering the full energy range can be generated after data parsing.

\begin{figure}[th]
    \centering
    \includegraphics[width=1.00\linewidth]{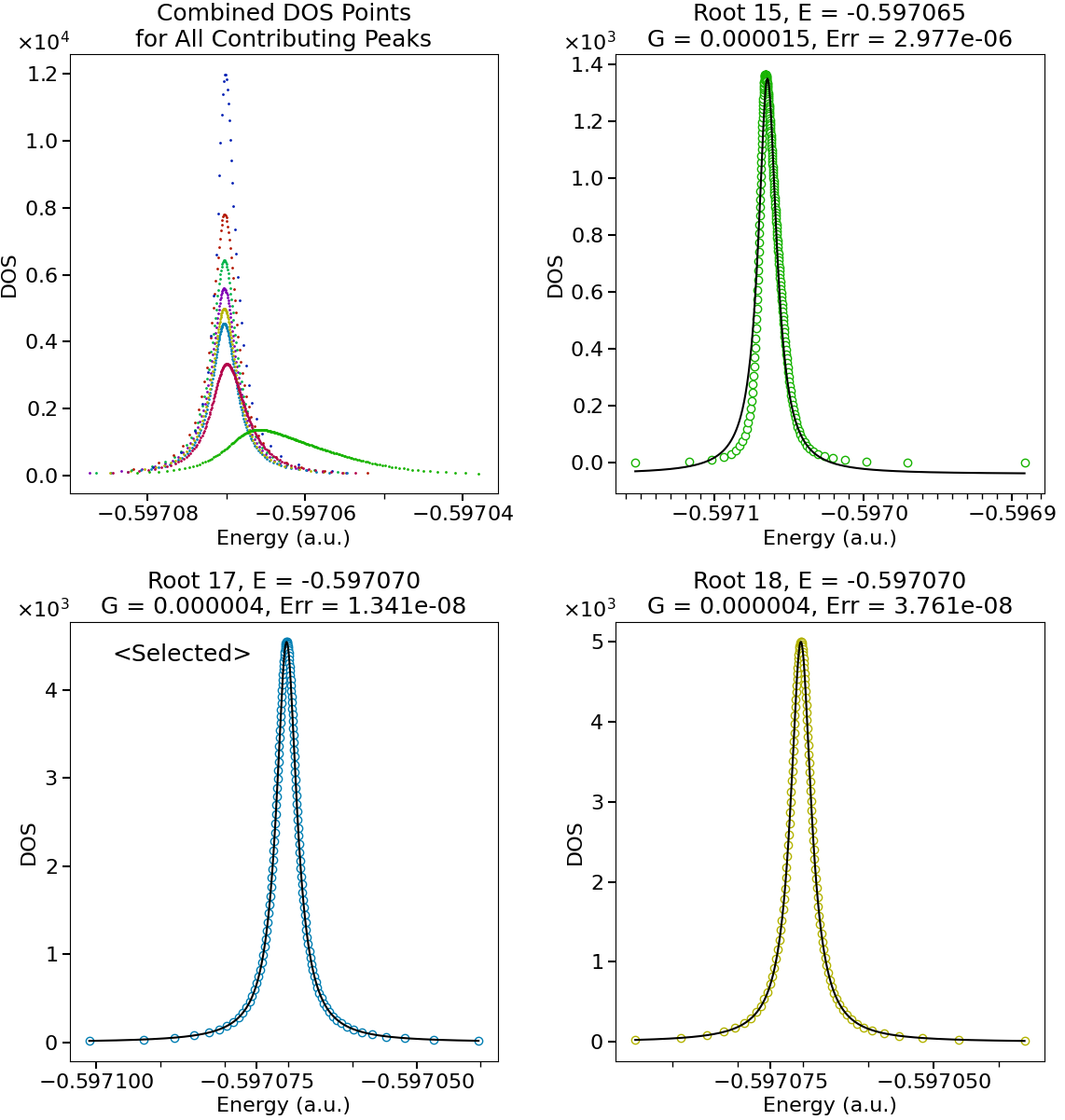}
    \caption{Subset of the \dosmax-generated grid plot displaying the DOS points (colored dots) and corresponding Lorentzian fits (black lines) of roots \textcolor[HTML]{08B849}{15}, \textcolor[HTML]{1789CC}{17}, and \textcolor[HTML]{ABAE0B}{18} for the second RS of ${}^1\text{P}^{\text{o}}$ ${}^{\infty}\text{He}$. Each peak fit plot is annotated with the root number, peak position $E \equiv E_r$ and width $G \equiv \Gamma$, as well as the relative SSR per point $\text{Err} \equiv \chi_{r}$. The top-left subplot displays the DOS data points of all contributing roots together, providing an overview of their relative positions and heights. It demonstrates that the DOS peaks \textcolor[HTML]{08B849}{15} and \textcolor[HTML]{B80808}{16} (fit not shown here) do not have fully Lorentzian character, in contrast to the remaining peaks \textcolor[HTML]{1789CC}{17} and \textcolor[HTML]{ABAE0B}{18}. The DOS peak of root \textcolor[HTML]{1789CC}{17} has been selected as the best fit, as indicated by a label \texttt{<Selected>} and the smallest fit error.}
    \label{fig:nice_fit_grid}
\end{figure}

For each resonance \texttt{i}, its DOS peaks and associated Lorentzian fits (if available) can be visualized with the command \texttt{grid i}. For the $2^\text{nd}$ RS in our example, \texttt{grid 2} produces the plot shown in Fig.~\ref{fig:nice_fit_grid}, confirming that a suitable representative peak has been selected for this RS.

\subsection{Manual Refinement}
\begin{figure*}[th]
    \centering
    \begin{tikzpicture}
        \node[anchor=south west, inner sep=0] (image) at (0,0) {\includegraphics[width=0.96\linewidth, trim=16 12 0 0, clip]{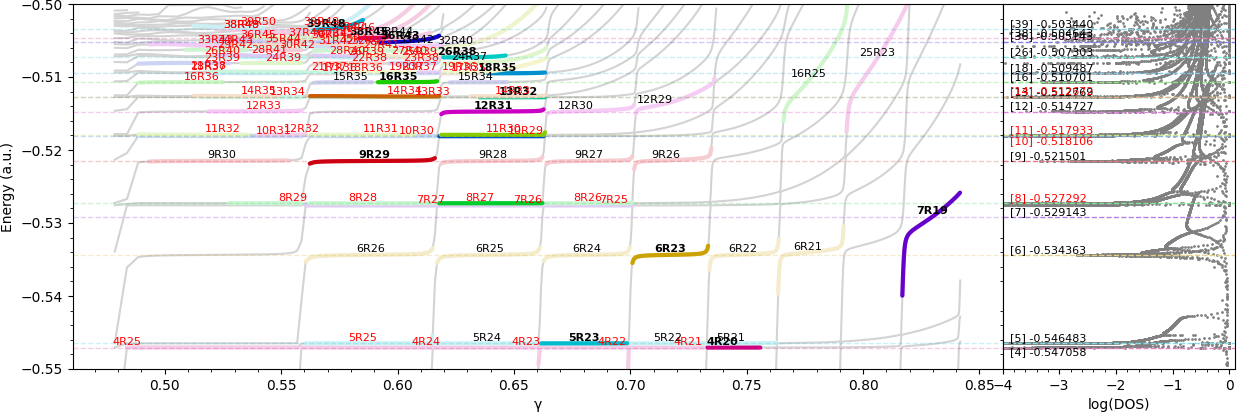}};
        \begin{scope}[x={(image.south east)}, y={(image.north west)}]
            \node[anchor=north] at (0.4, 0) {\scriptsize$\gamma$};
            \node[anchor=north] at (0.9, 0) {\scriptsize$\log_{10}(\rho)$};
            \node[rotate=90, anchor=south] at (0, 0.55) {\scriptsize$E$ (a.u.)};
        \end{scope}
    \end{tikzpicture}
    \caption{\dosmax-generated resonance overview plot corresponding to ${}^{1}\text{P}^{\text{o}}$ ${}^{\infty\!}\text{He}$ in the energy range from $-0.55$ to $-0.5$ a.u., consisting of a stabilization diagram with highlighted plateaus and detected RSs (on the left), and a logarithmic scatter plot of the  DOS distributions, indicating resonances as clear peaks (on the right). For each resonance, the representative plateau is indicated in a darker shade, and labeled in bold. Descending plateaus are labeled in red, as are resonances that have a descending plateau as their representative.}
    \label{fig:resonance_overview_not_nice}
\end{figure*}
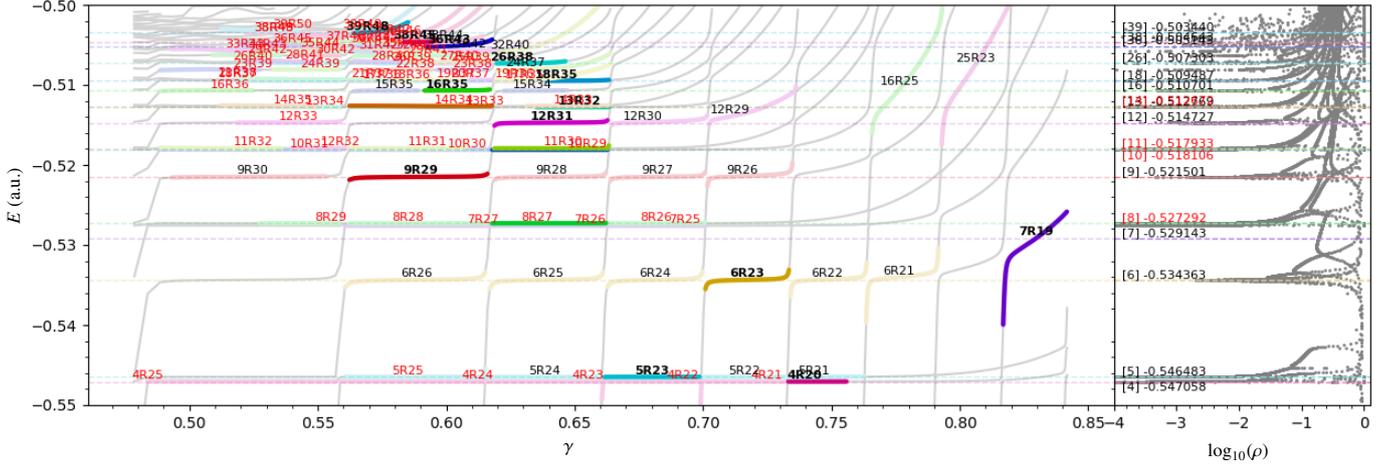
\begin{figure*}[th]
    \centering
    \begin{tikzpicture}
        \node[anchor=south west, inner sep=0] (image) at (0,0) {\includegraphics[width=0.96\linewidth, trim=16 12 0 0, clip]{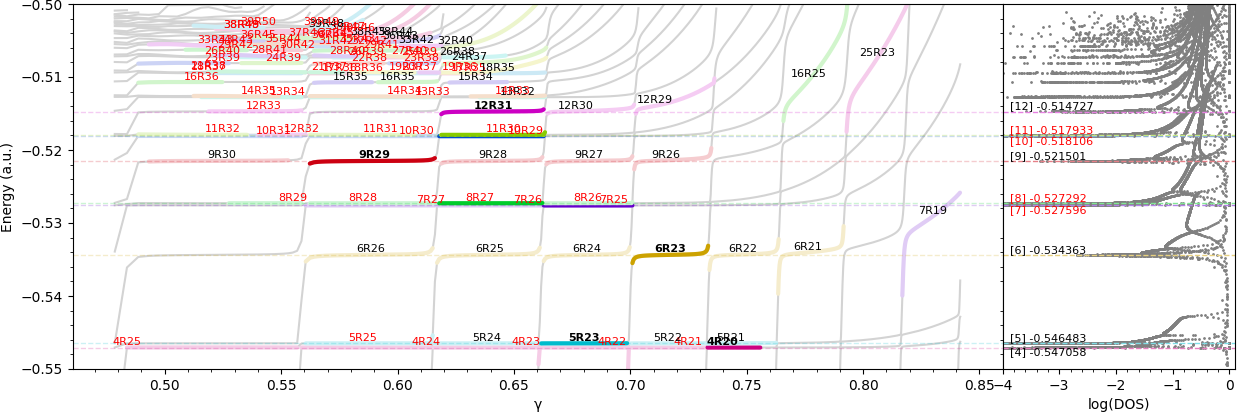}};
        \begin{scope}[x={(image.south east)}, y={(image.north west)}]
            \node[anchor=north] at (0.4, 0) {\scriptsize$\gamma$};
            \node[anchor=north] at (0.9, 0) {\scriptsize$\log_{10}(\rho)$};
            \node[rotate=90, anchor=south] at (0, 0.55) {\scriptsize$E$ (a.u.)};
        \end{scope}
    \end{tikzpicture}
    \caption{Resonance overview plot corresponding to ${}^{1}\text{P}^{\text{o}}$ ${}^{\infty\!}\text{He}$ in the energy range from $-0.55$ to $-0.5$ a.u. after manual adjustments achieved by entering \texttt{7R25 13+R} in \dosmax. The DOS clustering plot on the right side shows at a glance that all reliable RS energies are now selected correctly.}
    \label{fig:cleaned_plot}
\end{figure*}

As one approaches a HIT from below, RSs become increasingly dense, and some are no longer well-described by the available basis set. This effect is apparent in the upper portion of the resonance overview plot in Fig.~\ref{fig:resonance_overview_not_nice}.

This inadequate stabilization of RSs in such regions makes automatic detection challenging, often necessitating human intervention. An example of incorrect automatic identification of a representative for an RS is shown in Fig.~\ref{fig:resonance_overview_not_nice}, where the plateau \texttt{7R19} is incorrectly assigned to RS~7, and even erroneously recognized as its representative, because all other plateaus in this RS have slightly descending character indicated in Fig.~\ref{fig:resonance_overview_not_nice} by red labels (compare also the Subsection {\bf{Descending energy plateaus}} in \ref{subsec:peak-fitting}).
Since the misplaced plateau \texttt{7R19} is the only one in RS 7 with a valid DOS fit, it is then falsely selected as its best fit, resulting in completely wrong  automatic detection of parameters for this resonance. Human intervention in this case could look as follows. An execution of the command \texttt{grid 7} and inspection of the resulting plots reveal that the energies of the descending plateaus are very close: $-0.527596$, $-0.527596$, and $-0.527595$~a.u.~for the plateaus \texttt{7R25}, \texttt{7R26}, and \texttt{7R27}, respectively. We may therefore conclude that RS 7 is well represented by one of them (e.g., \texttt{7R25}) and its energy estimate obtained in this way is quite accurate; unfortunately, the stabilization method with DOS computed using Eq.~(\ref{eq:DOS}) is not capable of estimating the width of this resonance with the basis set used in the current example. Such descending sections frequently occur for RSs located in close proximity of the HIT, particularly at lower $\gamma$ values, due to the incompleteness of the basis set. Interestingly, such descending segments may also indicate fluorescence-active RSs, a possibility which will be discussed further in Section~\ref{results}.

At energies slightly below the HITs, non-monotonic energy behavior of the roots becomes more and more prevalent, introducing spurious features in the SDs, characterized by sharply descending lines or wave patterns without ever properly forming plateaus.  This limitation of the stabilization method is clearly visible in the upper portion of Fig.~\ref{fig:resonance_overview_not_nice}, where all the resonances from RS 13 up should not be deemed reliable. In practical applications, it is the user’s responsibility to select the energy trust range that yields reliable results when using \dosmax. A useful approach to determine such trust range could be based on a comparison of \dosmax results with analogous results computed with a slightly larger basis set: Resonances characterized by almost identical parameters can be assumed to be stabilized.

Various useful commands are available in \dosmax to help the user in manipulating and adjusting the results for a given resonance \textit{i}:
\begin{itemize}
    \item \texttt{iRj} selects the DOS peak of root $j$ as the resonance's representative.
    \item \texttt{iR} (without a root number) completely removes the resonance from consideration.
    \item \texttt{i+R} removes all resonances starting from \textit{i} up to the next threshold.
    \item \texttt{grid i} displays all DOS peaks of the resonance, together with their Lorentzian fits.
\end{itemize}
These commands can be combined. For example, to select root 25 as a representative for RS 7, and to deactivate the unreliable resonances RS 13 and above, the user should input \texttt{7R25 13+R}. Changing the resonance selection causes the associated resonance overview plot to be immediately rerendered: The resonance overview plots before and after invoking this command are shown in Figs.~\ref{fig:resonance_overview_not_nice} and~\ref{fig:cleaned_plot}, respectively. The resonances and plateaus selected in Fig.~\ref{fig:cleaned_plot} are exactly those listed in the first two columns of Table~\ref{tab:rsp_p-state}, resonance energies obtained from descending plateaus are presented there without the corresponding $\Gamma$ value.

\begin{figure}[th]
    \centering
    \includegraphics[width=1.00\linewidth]{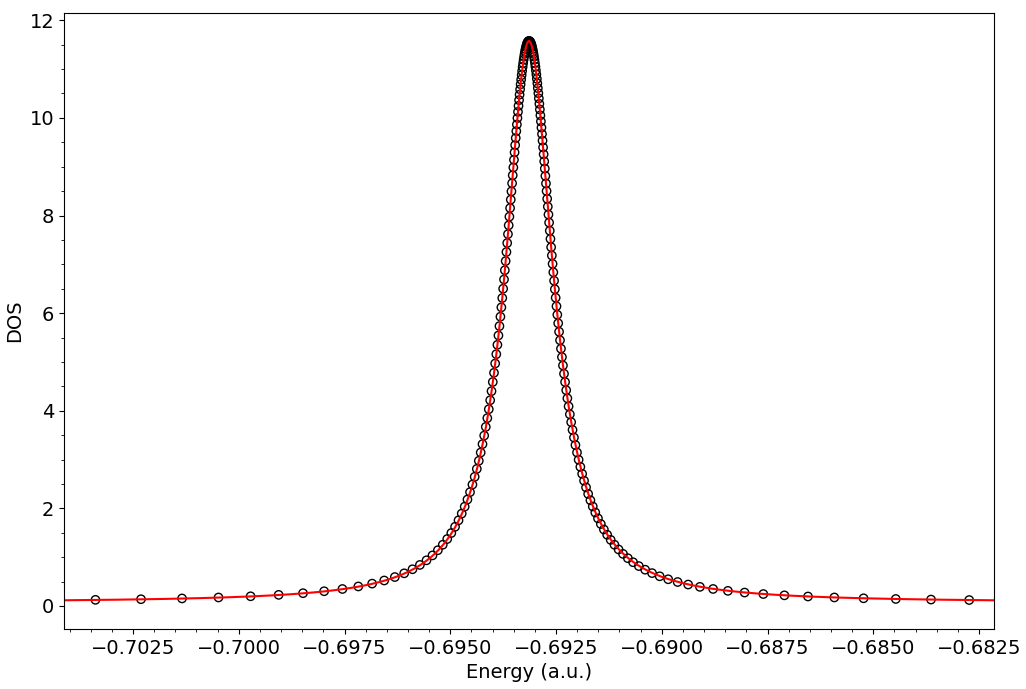}
    \caption{Lorentzian fit of the DOS peak selected to represent the first resonance state of ${}^1\text{P}^{\textnormal{o}}$ ${}^{\infty}\text{He}$, with the resonance parameters $E_r = -0.693139$, $\Gamma = 0.00137397$, an error of $\chi_r = 3.92715559 \times 10^{-6}$, occurring at $\gamma=0.8182$.}
    \label{fig:fit_1}
\end{figure}

Once the user is satisfied with the selection of all resonances under a given threshold, they can proceed to the next threshold, revisit the previous one, or finalize the adjustments, discarding all unreviewed thresholds. The final resonance parameters are  written to the file \texttt{results.txt}. If desired, the corresponding plots of the Lorentzian fits of the representative DOS peak of each resonance, such as the one shown in Fig.~\ref{fig:fit_1}, can be generated as \texttt{.png} files.

\vspace{0.5cm}
\section{Computed Resonance Parameters}\label{results}

We have used \dosmax to obtain the parameters $E_r$ and $\Gamma$ for the ${}^{1}\textnormal{S}^{\textnormal{e}}$, ${}^{3}\textnormal{S}^{\textnormal{e}}$, ${}^{1}\textnormal{P}^{\textnormal{o}}$,  and ${}^{3}\textnormal{P}^{\textnormal{o}}$ resonance states of helium with a finite ($m_{\alpha}=7294.299\,541\,71$~a.u. \cite{codata2022}) and infinite mass ($m_{\alpha}=\infty$)  below the He$^+(2s)$ ionization threshold. For each of 1000 distinct values of the basis set parameter $\gamma$, ranging from $0.4729$ to $0.8458$, the energy spectrum  has been computed by solving the GEP in Eq.~\eqref{gep}. For each value of $\gamma$, the basis set containing 720 explicitly correlated Hylleraas-type basis functions  has been constructed according to Eq.~\eqref{rwf} with $n=9$, $\texttt{w}=3$, and $\lambda_{0}=4.0$ for the $\textnormal{S}^{\textnormal{e}}$ states and with $n=9$, $\texttt{w}=2$, and $\lambda_{0}=4.0$ for the $\textnormal{P}^{\textnormal{o}}$ states. The  resonance parameters extracted from the resulting data by \dosmax are tabulated in Tables~\ref{tab:rsp_s-state} and~\ref{tab:rsp_p-state}. The RSs are listed in ascending energy order.

A total of 25 RSs with $L=0$ (15 ${}^{1}\textnormal{S}^{\textnormal{e}}$  and 10 ${}^{3}\textnormal{S}^{\textnormal{e}}$ states) of helium  below the He$^+(2s)$ threshold have been identified by \dosmax and are reported here. The  data for the ${}^{1}\textnormal{S}^{\textnormal{e}}$ and ${}^{3}\textnormal{S}^{\textnormal{e}}$ states computed with $m_{\alpha}=\infty$ are in excellent agreement with the values estimated by B\"{u}rgers \textit{et al.}\ \cite{Burgers1995} using the complex rotation technique with a $24\,497$-term Laguerre polynomial basis in perimetric coordinates. In contrast, our basis set of explicitly correlated Hylleraas functions is considerably smaller. Reference data for RSs of helium with a finite nuclear mass are scarce. Our results are in good agreement with the results published recently by Sil \textit{et al.}\ \cite{Sil2024}, who reported the first four resonances of the $^{1}$S$^e$ state of helium using the stabilization method with the mass-polarization term treated as a first-order time-independent perturbation. To the best of our knowledge, the parameters for the remaining 11  $^{1}\text{S}^e$ and all 10 $^{3}$S$^e$ resonance states of helium computed with finite nuclear mass have never been reported before. The differences between the resonance parameters computed with $m_{\alpha}=7294.299\,541\,71$~a.u. and $m_{\alpha}=\infty$ seem to be a few times smaller than the analogous differences observed for the bound states. Nakashima and Nakatsuji \cite{Nakashima2007,Nakashima2008} reported the ground state energy of $-2.903\,304\,557$~a.u.\ for a slightly less accurate $\alpha$ particle mass $m_{\alpha}=7294.299\,536\,5$~a.u.\ \cite{codata2006} and $-2.903\,724\,377$~a.u.\ for $m_{\alpha}=\infty$, which gives the energy difference of 4.2$\times$10$^{-4}$~a.u., while in our case the corresponding energy differences vary from 0.6$\times$10$^{-4}$~a.u.\ to 1.2$\times$10$^{-4}$~a.u.

Similarly, a total of 21 RSs with $L=1$ (12 ${}^{1}\textnormal{P}^{\textnormal{o}}$  and 9 ${}^{3}\textnormal{P}^{\textnormal{o}}$ states) of helium  below the He$^+(2s)$ threshold have been identified by \dosmax and are reported here. The  data for the ${}^{1}\textnormal{P}^{\textnormal{o}}$  and ${}^{3}\textnormal{P}^{\textnormal{o}}$ states computed with $m_{\alpha}=\infty$ and shown in Table~\ref{tab:rsp_p-state} are in good agreement with available literature values \cite{Lasso2013,Chen1997,kar2006}. Table~\ref{tab:rsp_p-state} shows that RS~4 for ${}^{1}\textnormal{P}^{\textnormal{o}}$  and ${}^{3}\textnormal{P}^{\textnormal{o}}$  exhibit considerably smaller widths than other RSs, which allows us to categorize them as fluorescence-active RSs according to symmetry criteria proposed by Saha \textit{et al.}\ \cite{saha2011a}. Such states  are more likely to decay via the fluorescence channel than via the autoionization channel. Similar characterization is most likely true also for the ${}^{1}\textnormal{P}^{\textnormal{o}}$  and ${}^{3}\textnormal{P}^{\textnormal{o}}$ states RS~7 and RS~10, characterized by very flat but non-monotonic energy plateaus; here, the stabilization method as implemented in \dosmax does not allow us to determine the resonance widths with the basis of 720 explicitly correlated Hylleraas-type basis functions obtained with  $n=9$, $\texttt{w}=2$, and $\lambda_{0}=4.0$. This problem can be readily alleviated by employing a larger basis set. Changing  $\texttt{w} = 2$ to $\texttt{w} = 3$ increases the basis set size to 1440 basis functions and resolves the nonmonotonicity problem for these ${}^{3}\textnormal{P}^{\textnormal{o}}$ RSs computed with $m_{\alpha}=\infty$, yielding the resonance energies $-0.54884$, $-0.52864$, $-0.51871$~a.u.\ and resonance widths $1.25\times10^{-8}$, $5.64\times10^{-9}$, $1.69\times10^{-9}$~a.u.\ for RS~4, RS~7, RS~10, respectively.  Moreover, with this extended basis,  it is possible to determine also the resonance parameters corresponding to the high-lying $^3\text{P}^{\text{o}}$ states RS~11 and RS~12; we get  $E_{r}=-0.51711$ a.u.\ and  $\Gamma=8.15\times10^{-6}$~a.u.\ for RS~11 and $E_{r}=-0.51608$ a.u.\ and $\Gamma=2.36\times10^{-7}$~a.u.\ for RS~12. All these values are in good agreement with those reported earlier by Chen \cite{Chen1997}. Files used in the analysis of the $^3\text{P}^{\text{o}}$ state of $^\infty\text{He}$ with the extended basis are available in the \texttt{results} directory on \texttt{GitHub} \cite{Langner2025}.

\begin{scriptsize}
\begin{table*}
\caption{Resonance energies ($E_r$ in a.u.) and widths ($\Gamma$ in a.u.) of the ${}^{1}\textnormal{S}^{\textnormal{e}}$ and ${}^{3}\textnormal{S}^{\textnormal{e}}$ states of helium with infinite ($^{\infty}\text{He}$) and finite ($\text{He}$) nuclear mass located below the He$^+(2s)$ ionization threshold. The symbol $\texttt{a}(-\texttt{b})$ stands for $\texttt{a}\times 10^{-\texttt{b}}$.}\label{tab:rsp_s-state}
\vspace{0.5cm}
\begin{tabular}{cllllllllllll}
\hline\hline\\
\vspace{-0.75cm}\\
&&\multicolumn{5}{c}{$^1$S$^e$}&&\multicolumn{5}{c}{$^3$S$^e$}\\
\cline{3-7}\cline{9-13}\\
\vspace{-0.75cm}\\
&&\multicolumn{2}{c}{$^{\infty}\text{He}$}&&\multicolumn{2}{c}{$\text{He}$}&&\multicolumn{2}{c}{$^{\infty}\text{He}$}&&\multicolumn{2}{c}{$\text{He}$}\\
\cline{3-4}\cline{6-7}\cline{9-10}\cline{12-13}\\
\vspace{-0.75cm}\\
State&&\multicolumn{1}{c}{$-E_r$}&\multicolumn{1}{c}{$\Gamma$}&&\multicolumn{1}{c}{$-E_r$}&\multicolumn{1}{c}{$\Gamma$}&&\multicolumn{1}{c}{$-E_r$}&\multicolumn{1}{c}{$\Gamma$}&&\multicolumn{1}{c}{$-E_r$}&\multicolumn{1}{c}{$\Gamma$}\\
\hline\\
1&&0.77787\phantom{$^a$}&4.54(-3)\phantom{$^a$}&&0.77777\phantom{$^a$}&4.54(-3)\phantom{$^a$}&& 0.60257\phantom{$^a$} &  6.70(-6)\phantom{$^a$}  & & 0.60249\phantom{$^a$} & 6.69(-6)\phantom{$^a$} \\
 && 0.77787$^a$ & 4.54(-3)$^a$ &&0.77749$^b$&\phantom{4.54(-3)}&& 0.60257$^a$ &  6.63(-6)$^a$ &&\phantom{0.60249}&\phantom{6.69(-6)} \\[3pt]
2&& 0.62193\phantom{ $^a$} & 2.16(-4)\phantom{ $^a$} &&0.62183\phantom{ $^a$}&2.16(-4)\phantom{ $^a$}&& 0.55975\phantom{ $^a$} & 2.61(-7)\phantom{ $^a$} && 0.55967\phantom{ $^a$} & 2.61(-7)\phantom{ $^a$}\\
 && 0.62193$^a$ & 2.16(-4)$^a$ && 0.62183$^b$&&& 0.55975$^a$ & 2.60(-7)$^a$ \\[3pt]
3&& 0.58989\phantom{ $^a$} & 1.37(-3)\phantom{ $^a$} &&0.58982\phantom{ $^a$}&1.37(-3)\phantom{ $^a$}&& 0.54884\phantom{ $^a$} & 3.19(-6)\phantom{ $^a$}&& 0.54876\phantom{ $^a$} & 3.19(-6)\phantom{ $^a$}\\
 && 0.58989$^a$ & 1.36(-3)$^a$ && 0.58971$^b$&&& 0.54884$^a$ & 3.09(-6)$^a$\\[3pt]
4&& 0.54809\phantom{ $^a$} & 7.47(-5)\phantom{ $^a$} &&0.54801\phantom{ $^a$}&7.54(-5)\phantom{ $^a$}&& 0.53251\phantom{ $^a$} & 1.43(-7)\phantom{ $^a$} && 0.53243\phantom{ $^a$} & 1.43(-7)\phantom{ $^a$}\\
 && 0.54809$^a$ & 7.48(-5)$^a$ && 0.54800$^b$&&& 0.53251$^a$ & 1.44(-7)$^a$ \\[3pt]
5&& 0.54488\phantom{ $^a$} & 4.98(-4)\phantom{ $^a$} &&0.54481\phantom{ $^a$}&4.98(-4)\phantom{ $^a$}&& 0.52841\phantom{ $^a$} & 1.54(-6)\phantom{ $^a$} &&  0.52834\phantom{ $^a$} & 1.54(-6)\phantom{ $^a$} \\
 && 0.54488$^a$ & 4.92(-4)$^a$ &&&&& 0.52841$^a$ & 1.54(-6)$^a$ \\[3pt]
6&& 0.52772\phantom{ $^a$} & 4.61(-5)\phantom{ $^a$} &&0.52764\phantom{ $^a$}&4.74(-5)\phantom{ $^a$}&& 0.52055\phantom{ $^a$} & 8.13(-8)\phantom{ $^a$}&& 0.52048\phantom{ $^a$} & 8.11(-8)\phantom{ $^a$}\\
 && 0.52772$^a$ & 4.62(-5)$^a$ &&&&& 0.52055$^a$ & 8.20(-8)$^a$ \\[3pt]
7&& 0.52669\phantom{ $^a$} & 2.17(-4)\phantom{ $^a$} &&0.52662\phantom{ $^a$}&2.26(-4)\phantom{ $^a$}&& 0.51855\phantom{ $^a$} & 8.57(-7)\phantom{ $^a$} && 0.51848\phantom{ $^a$} & 8.57(-7)\phantom{ $^a$}\\
 && 0.52669$^a$ & 2.19(-4)$^a$ &&&&& 0.51855$^a$ & 8.56(-7)$^a$ \\[3pt]
8&& 0.51810\phantom{ $^a$} & 2.97(-5)\phantom{ $^a$} &&0.51803\phantom{ $^a$}&2.97(-5)\phantom{ $^a$}&& 0.51418\phantom{ $^a$} & 5.73(-8)\phantom{ $^a$} && 0.51411\phantom{ $^a$} & 1.49(-8)\phantom{ $^a$}\\
 && 0.51810$^a$ & 2.98(-5)$^a$ &&&&& 0.51418$^a$ & 5.00(-8)$^a$ \\[3pt]
9&& 0.51764\phantom{ $^a$} & 1.13(-4)\phantom{ $^a$} &&0.51757\phantom{ $^a$}&1.13(-4)\phantom{ $^a$}&& 0.51305\phantom{ $^a$} & 4.72(-7)\phantom{ $^a$} && 0.51298\phantom{ $^a$} & 4.74(-7)\phantom{ $^a$}\\
 && 0.51764$^a$ & 1.14(-4)$^a$ &&&&& 0.51305$^a$ & 5.20(-7)$^a$ \\[3pt]
10&& 0.51276\phantom{ $^a$} & 1.98(-5)\phantom{ $^a$} &&0.51269\phantom{ $^a$}&1.98(-5)\phantom{ $^a$}&& 0.51038\phantom{ $^a$} & 5.45(-7)\phantom{ $^a$} && 0.51031\phantom{ $^a$} & 5.31(-7)\phantom{ $^a$}\\
  && 0.51276$^a$ & 1.99(-5)$^a$ &&&&& 0.51038$^a$ & 3.20(-8)$^a$ \\[3pt]
11&& 0.51252\phantom{ $^a$} & 6.64(-5)\phantom{ $^a$} && 0.51244\phantom{ $^a$}&6.55(-5)\phantom{ $^a$}&& & &&0.50960\phantom{ $^a$}&4.23(-6)\phantom{ $^a$}\\
  && 0.51251$^a$ & 6.60(-5)$^a$ \\[3pt]
12&& 0.50948\phantom{ $^a$} & 1.38(-5)\phantom{ $^a$} && 0.50941\phantom{ $^a$}&1.40(-5)\phantom{ $^a$}&&\\
  && 0.50948$^a$ & 1.38(-5)$^a$ \\[3pt]
13&& 0.50933\phantom{ $^a$} & 4.19(-5)\phantom{ $^a$} && 0.50926\phantom{ $^a$}&4.25(-5)\phantom{ $^a$}&&\\
  && 0.50933$^a$ & 4.16(-5)$^a$ \\[3pt]
14&& 0.50732\phantom{ $^a$} & 1.37(-6)\phantom{ $^a$} && 0.50726\phantom{ $^a$}&9.81(-6)\phantom{ $^a$}&&\\
  && 0.50732$^a$ & 9.92(-6)$^a$ \\[3pt]
15&& 0.50722\phantom{ $^a$} & 3.44(-5)\phantom{ $^a$} && 0.50716\phantom{ $^a$}&3.45(-5)\phantom{ $^a$}&&\\
  && 0.50722$^a$ & 2.79(-5)$^a$ \\[3pt]
\hline\hline
\end{tabular}
\begin{tabular}{l} 
\\
$^a\phantom{b}$A.\ B\"{u}rgers \textit{et al.}: Complex rotation technique \cite{Burgers1995};\\
$^b\phantom{a}$A.\ N.\ Sil \textit{et al.}: Stabilization method \cite{Sil2024}.
\end{tabular}
\end{table*}
\end{scriptsize}

\begin{scriptsize}
\begin{table*}
\caption{Resonance energies ($E_r$ in a.u.) and widths ($\Gamma$ in a.u.) of the ${}^{1}\textnormal{P}^{\textnormal{o}}$ and ${}^{3}\textnormal{P}^{\textnormal{o}}$ states of helium with infinite ($^{\infty}\text{He}$) and finite ($\text{He}$) nuclear mass located below the He$^+(2s)$ ionization threshold. The symbol $\texttt{a}(-\texttt{b})$ stands for $\texttt{a}\times 10^{-\texttt{b}}$.}\label{tab:rsp_p-state}
\vspace{0.5cm}
\begin{tabular}{cllllllllllll}
\hline\hline\\
\vspace{-0.75cm}\\
&&\multicolumn{5}{c}{$^1$P$^\textnormal{o}$}&&\multicolumn{5}{c}{$^3$P$^\textnormal{o}$}\\
\cline{3-7}\cline{9-13}\\
\vspace{-0.75cm}\\
&&\multicolumn{2}{c}{$^{\infty}\text{He}$}&&\multicolumn{2}{c}{$\text{He}$}&&\multicolumn{2}{c}{$^{\infty}\text{He}$}&&\multicolumn{2}{c}{$\text{He}$}\\
\cline{3-4}\cline{6-7}\cline{9-10}\cline{12-13}\\
\vspace{-0.75cm}\\
State&&\multicolumn{1}{c}{$-E_r$}&\multicolumn{1}{c}{$\Gamma$}&&\multicolumn{1}{c}{$-E_r$}&\multicolumn{1}{c}{$\Gamma$}&&\multicolumn{1}{c}{$-E_r$}&\multicolumn{1}{c}{$\Gamma$}&&\multicolumn{1}{c}{$-E_r$}&\multicolumn{1}{c}{$\Gamma$}\\
\hline\\
1&&0.69314\phantom{$^a$}&1.37(-3)\phantom{ $^a$}&&0.69304\phantom{ $^a$}&1.38(-3)\phantom{ $^a$}&& 0.76049\phantom{$^a$} &  2.99(-4)\phantom{$^a$}  & & 0.76039\phantom{ $^a$} & 2.98(-4)\phantom{ $^a$} \\
 && 0.69293$^a$ & 1.30(-3)$^a$ &&  & && 0.76050$^a$ &3.12(-4)$^a$\\
 && 0.69307$^b$ & 1.37(-3)$^b$ &&  & && 0.76049$^b$ &2.99(-4)$^b$\\
 && 0.69313$^c$ & 1.37(-3)$^c$ &&  & && 0.76049$^c$ &3.00(-4)$^c$\\[3pt]
2 &&0.59707\phantom{$^a$}&3.86(-6)\phantom{$^a$}&&0.59699\phantom{ $^a$}&3.85(-6)\phantom{ $^a$}&&0.58467\phantom{$^a$}&8.66(-5)\phantom{$^a$}&&0.58459\phantom{$^a$}&8.63(-5)\phantom{$^a$}\\
  &&0.59707$^a$ &3.85(-6)$^a$   &&  & && 0.58468$^a$ &8.56(-5)$^a$\\
  &&0.59707$^b$ &3.84(-6)$^b$   &&  & && 0.58467$^b$ &8.24(-5)$^b$\\
  &&0.59707$^c$ &3.88(-6)$^c$   &&  & && 0.58467$^c$ &8.25(-5)$^c$\\[3pt]
3 &&0.56408\phantom{$^a$}&3.08(-4)\phantom{$^a$}&&0.56400\phantom{ $^a$}&3.08(-4)\phantom{ $^a$}&&0.57902\phantom{$^a$}&1.90(-6)\phantom{$^a$}&&0.57894\phantom{$^a$}&1.89(-6)\phantom{$^a$}\\
  &&0.56404$^a$ & 2.85(-4)$^a$ &&  & && 0.57903$^a$ &1.80(-6)$^a$\\
  &&0.56407$^b$ & 2.99(-4)$^b$ &&  & && 0.57903$^b$ &1.85(-6)$^b$\\
  &&0.56408$^c$ & 3.02(-4)$^c$ &&  & && 0.57903$^c$ &1.89(-6)$^c$\\[3pt]
4 &&0.54706\phantom{$^a$}&3.36(-8)\phantom{$^a$}&&0.54698\phantom{ $^a$}&3.39(-8)\phantom{$^a$}&&0.54882\phantom{$^a$}&\phantom{ $^a$}&&0.54874\phantom{$^a$}&\phantom{ $^a$}\\
  &&0.54707$^a$ &1.82(-8)$^a$  &&  & && 0.54884$^a$ &2.74(-8)$^a$\\
  &&0.54708$^b$ &1.50(-8)$^b$  &&  & && 0.54884$^b$ &1.30(-8)$^b$\\
  &&0.54709$^c$ &              &&  & && 0.54884$^c$ &            \\[3pt]
5 &&0.54648\phantom{$^a$}&2.00(-6)\phantom{$^a$}&&0.54641\phantom{$^a$}&1.86(-6)\phantom{$^a$}&&0.54284\phantom{$^a$}&3.18(-5)\phantom{$^a$}&&0.54276\phantom{$^a$}&3.17(-5)\phantom{$^a$}\\
  &&0.54648$^a$ & 2.02(-6)$^a$ &&  & && 0.54283$^a$ &3.27(-5)$^a$\\
  &&0.54649$^b$ & 2.02(-6)$^b$ &&  & && 0.54284$^b$ &3.17(-5)$^b$\\
  &&0.54649$^c$ & 2.03(-6)$^c$ &&  & && 0.54284$^c$ &3.17(-5)$^c$\\[3pt]
6 &&0.53436\phantom{$^a$}&1.35(-4)\phantom{$^a$}&&0.53429\phantom{$^a$}&1.35(-4)\phantom{$^a$}&&0.53956\phantom{$^a$}&1.04(-6)\phantom{$^a$}&&0.53948\phantom{$^a$}&1.04(-6)\phantom{$^a$}\\
  &&0.53431$^a$ & 1.22(-4)$^a$ &&  & && 0.53953$^a$ &7.91(-7)$^a$\\
  &&0.53436$^b$ & 1.28(-4)$^b$ &&  & && 0.53956$^b$ &7.90(-7)$^b$\\
  &&0.53436$^c$ & 1.29(-4)$^c$ &&  & && 0.53956$^c$ &8.9(-7)$^c$\\[3pt]
7 &&0.52760\phantom{$^a$} &  &&0.52752\phantom{$^a$} & &&0.52862\phantom{$^a$}&\phantom{$^a$}&&0.52855\phantom{$^a$}&\\
  &&0.52757$^a$ &  &&  & && 0.52817$^a$ &\\
  &&0.52761$^b$ & 3.00(-9)$^b$ &&  & && 0.52864$^b$ &6.60(-9)$^b$\\
  &&0.52762$^c$ &              &&  & && 0.52864$^c$ &            \\[3pt]
8 &&0.52729\phantom{$^a$} &  && 0.52722\phantom{$^a$} & &&0.52571\phantom{$^a$}&1.63(-5)\phantom{$^a$}&&0.52564\phantom{$^a$}&1.63(-5)\phantom{$^a$}\\
  &&0.52730$^b$ & 9.88(-7)$^b$ &&  & && 0.52571$^b$ &1.51(-5)$^b$\\
  &&0.52730$^c$ &              &&  & && 0.52571$^c$ &\phantom{6.60(-9)$^b$}\\[3pt]
9 &&0.52150\phantom{$^a$} & 5.37(-5)\phantom{$^a$} &&0.52142\phantom{$^a$} & 5.38(-5)\phantom{$^a$} &&0.52395\phantom{$^a$}&2.51(-7)\phantom{$^a$}&&0.52387\phantom{$^a$}&2.45(-7)\phantom{$^a$}\\
  &&0.52150$^b$ & 6.44(-5)$^b$ &&  & && 0.52395$^b$ &4.10(-7)$^b$\\
  &&0.52149$^c$ &              &&  & && 0.52395$^c$ &            \\[3pt]
10 &&0.51811\phantom{$^a$} &  &&0.51803\phantom{$^a$} & &&\phantom{$^a$}&\phantom{$^a$}&&&\\
  &&0.51812$^b$ & \phantom{6.44(-5)$^b$} &&  & && 0.51871$^b$ &3.20(-9)$^b$\\[3pt]
11 &&0.51793\phantom{$^a$}&  && & &&\phantom{$^a$}&\phantom{$^a$}&&&\\
 &&0.51794$^b$ & 5.40(-7)$^b$ &&  & && 0.51711$^b$ &8.60(-6)$^b$\\[3pt]
12 &&0.51473\phantom{$^a$} & 6.68(-5) && & &&&\phantom{$^a$}&&&\\
 &&0.51473$^b$ & 3.71(-5)$^b$ &&  & && 0.51608$^b$ &2.30(-7)$^b$\\[3pt]
\hline\hline
\end{tabular}
\begin{tabular}{l} 
$^a$ A. F. Ord\'o\~nez-Lasso \textit{et al.}: Feshbach projection method \cite{Lasso2013}. \\
$^b$ M.-K. Chen : Saddle-point complex-rotation method \cite{Chen1997}.\\
$^c$ S. Kar and Y. K. Ho : Stabilization method \cite{kar2006}.
\end{tabular}
\end{table*}
\end{scriptsize}

\section{Conclusion} \label{conclusion}
We report here the open-source software \dosmax designed to fully automate the extraction of resonance parameters --- position and width --- from a data file containing energy eigenvalues of atomic and molecular systems computed using the stabilization method with a single basis set parameter $\gamma$. By systematically computing the DOS, identifying stabilization plateaus, and fitting Lorentzian curves, \dosmax provides an efficient and reliable method for automatic detection and analysis of RSs, eliminating the need for labor-intensive manual DOS peak selection and fitting. However, as with any implementation of the stabilization method, automatic peak identification becomes increasingly  challenging for resonance states located directly below ionization thresholds, where the plateaus display pathological $\gamma$-dependence for incomplete basis sets. In such cases, \dosmax supports interactive manual adjustments enabling the user to exclude incorrectly detected resonances and retain only physically meaningful results. 

We demonstrate the functionality of \dosmax by using it to identify the resonance parameters for the ${}^{1}\textnormal{S}^{\textnormal{e}}$, ${}^{3}\textnormal{S}^{\textnormal{e}}$, ${}^{1}\textnormal{P}^{\textnormal{o}}$, and ${}^{3}\textnormal{P}^{\textnormal{o}}$ doubly excited resonance states of helium. For each of 1000 distinct values of the basis set parameter $\gamma$, the energy spectrum  has been computed by solving the generalized eigenvalue problem in a basis containing 720 explicitly correlated Hylleraas-type basis functions. The resonance parameters obtained from this data using \dosmax\ --- automatically computed within 3~sec and manually postprocessed  over an additional 10~min --- show good agreement with available literature values. The resonance parameters for several doubly excited resonance states of helium with finite nuclear mass are reported here for the first time.

\subsection*{Acknowledgements}
JL, AS, JKS and HW acknowledge support from the National Science and Technology Council (NSTC), Taiwan under grant number NSTC 113-2639-M-A49-002-ASP, NSTC 113-2811-M-A49-531, NSTC 113-2113-M-A49-001 and NSTC 113-2113-M-A49-001, respectively. JKS acknowledges partial financial support from the Anusandhan National Research Foundation (ANRF) [erstwhile Science and Engineering Research Board (SERB)], Govt.\ of India, under file number CRG/2022/003547. 

\subsection*{Data Availability Statement}
All data files that support the findings of this study and the used code \dosmax is published on GitHub \cite{Langner2025}.

\bibliographystyle{elsarticle-num}
\bibliography{biblio}

\end{document}